\documentclass[a4paper]{jpconf}
\usepackage{graphicx}
\begin{document}
\title{Search for muoproduction of the $X(3872)$ at COMPASS}

\author{Alexey Guskov for the COMPASS collaboration}

\address{Joint Institute for Nuclear Research, Dubna, Russia}

\ead{avg@jinr.ru}

\begin{abstract}
 Exotic charmonium-like states have been observed  by various experiments over the last 15 years,
but their nature is still under discussion. Photo-(muo)production is a new promising instrument to study them. 
COMPASS, a fixed target experiment at CERN, analyzed the full set of the data collected with a
muon beam between 2002 and 2011, covering the range from 7 GeV to 19 GeV in the centre-of-mass energy
of the virtual photon-nucleon system.  A signal in the mass spectrum of $J/\psi\pi^+\pi^-$ with the statistical significance
 of 4.1 $\sigma$ was observed in the reaction $\mu^+~N \rightarrow \mu^+(J/\psi\pi^+\pi^- )\pi^{\pm} N'$. Its mass and 
width are consistent with those of the $X(3872)$. The shape of the $\pi^+\pi^-$ mass distribution from the observed decay
 into $J/\psi\pi^+\pi^-$  is different from previous observations for $X(3872)$. The observed signal  may be
 interpreted as  possible evidence of a new charmonium state $\widetilde{X}(3872)$. It could be associated with a neutral partner of 
$X(3872)$ with $C = -1$ predicted by a tetraquark model. 
\end{abstract}

\section{Introduction}

Over the last years a lot of new charmonium-like hadrons, so-called the XYZ states, at the mass range above 3.8 GeV/$c^2$ were discovered.
Several interpretations of the new states do exist: pure quarkonia, tetraquarks, hadronic molecules, hybrid mesons with a gluon content, etc. But at the moment many basic parameters of the XYZ states have not been determined yet. New experimental input is required
to distinguish between the models that provide different interpretations of the nature of exotic charmonia. The search for exotic charmonium-like states in exclusive photoproduction reactions was for the first time proposed in \cite{photo0,photo1,photo2}.

COMPASS, a fixed-target experiment at a secondary beam of SPS at CERN \cite{proposal, COMPASS1}, already searched for photo(lepto-)production of the state $Z^{\pm}_c(3900)$ in the charge-exchange reaction $\mu^+N \to \mu^+ Z^{\pm}(3900)_c N'$ using the experimental data obtained for positive muons of 160~GeV/$c$ (2002-2010) or 200~GeV/$c$ momentum (2011)  scattering off solid $^6$LiD  (2002-2004) or NH$_3$ targets (2006-2011) \cite{Zc}. The search for muoproduction of the $X(3872)$ in exclusive 
 reactions is  the continuation of these studies.

\section{Exclusive muoproduction of X(3872)}
The exotic hadron $X(3872)$ was discovered by the Belle collaboration in 2003 \cite{x3872belle}. Its mass is 3871.69 $\pm$ 0.17 MeV/$c^2$  that is very close to the $D^0\bar{D}^{*0}$ threshold. The decay width of this state has not been determined yet, only an upper limit for the natural width $\Gamma_{X(3872)}$ of about 1.2 MeV/$c^2$ (CL$=$90\%) exists. The quantum numbers $J^{PC}$ of the $X(3872)$ were determined by LHCb to be $1^{++}$ \cite{x3872LHCB1,x3872LHCB2}. Approximately equal probabilities to decay into $J/\psi3\pi$ and $J/\psi2\pi$  final states indicate large isospin symmetry breaking.

\begin{figure}
 \begin{minipage}{18pc}
   \includegraphics[width=200px]{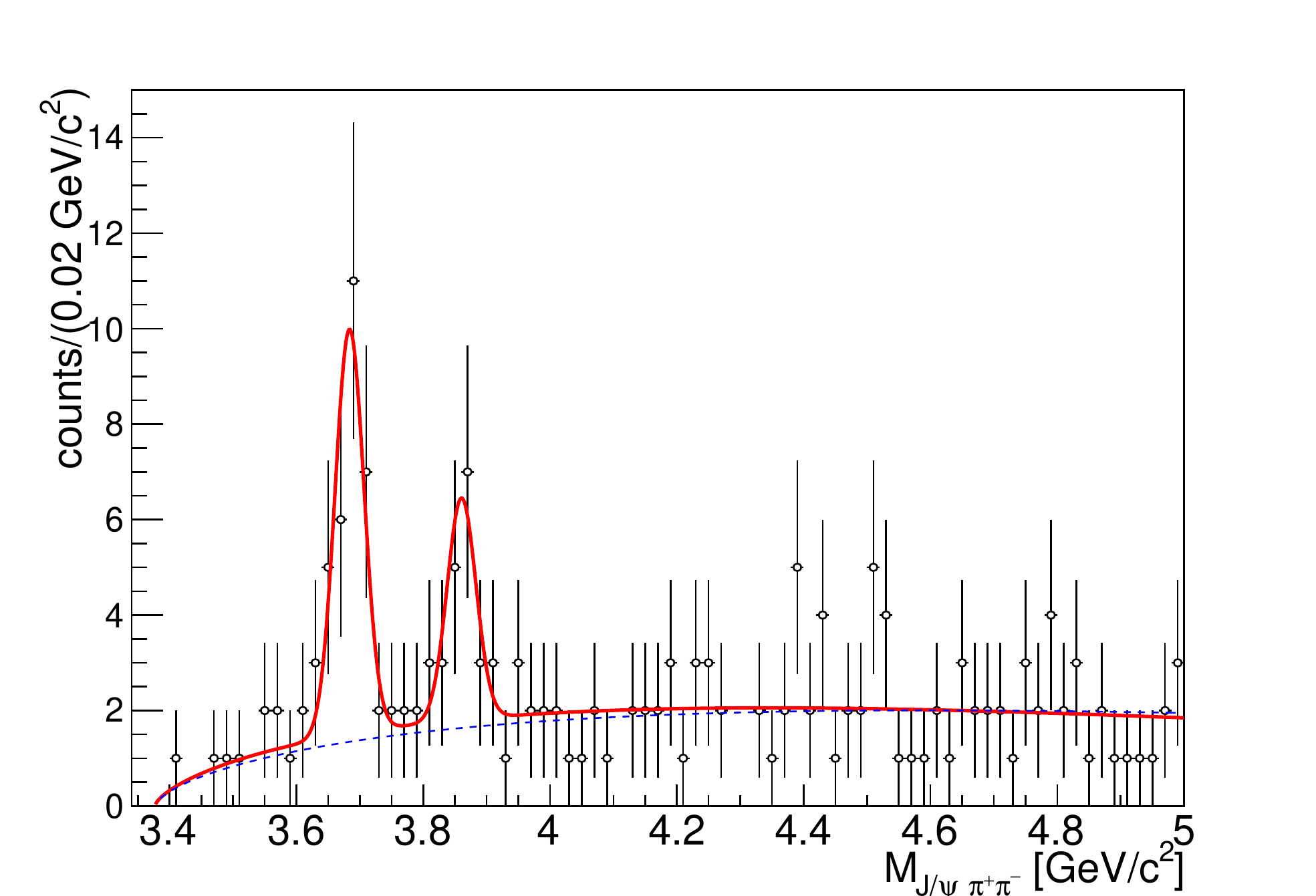}  
     \caption{\label{fig:x3872}
 The $J/\psi\pi^+\pi^-$ invariant mass distribution for the final state in the reaction $\mu^+ N \to \mu^+ J/\psi\pi^+\pi^-\pi^{\pm} N'$.}
\end{minipage}\hspace{2pc}%
\begin{minipage}{18pc}
   \includegraphics[width=200px]{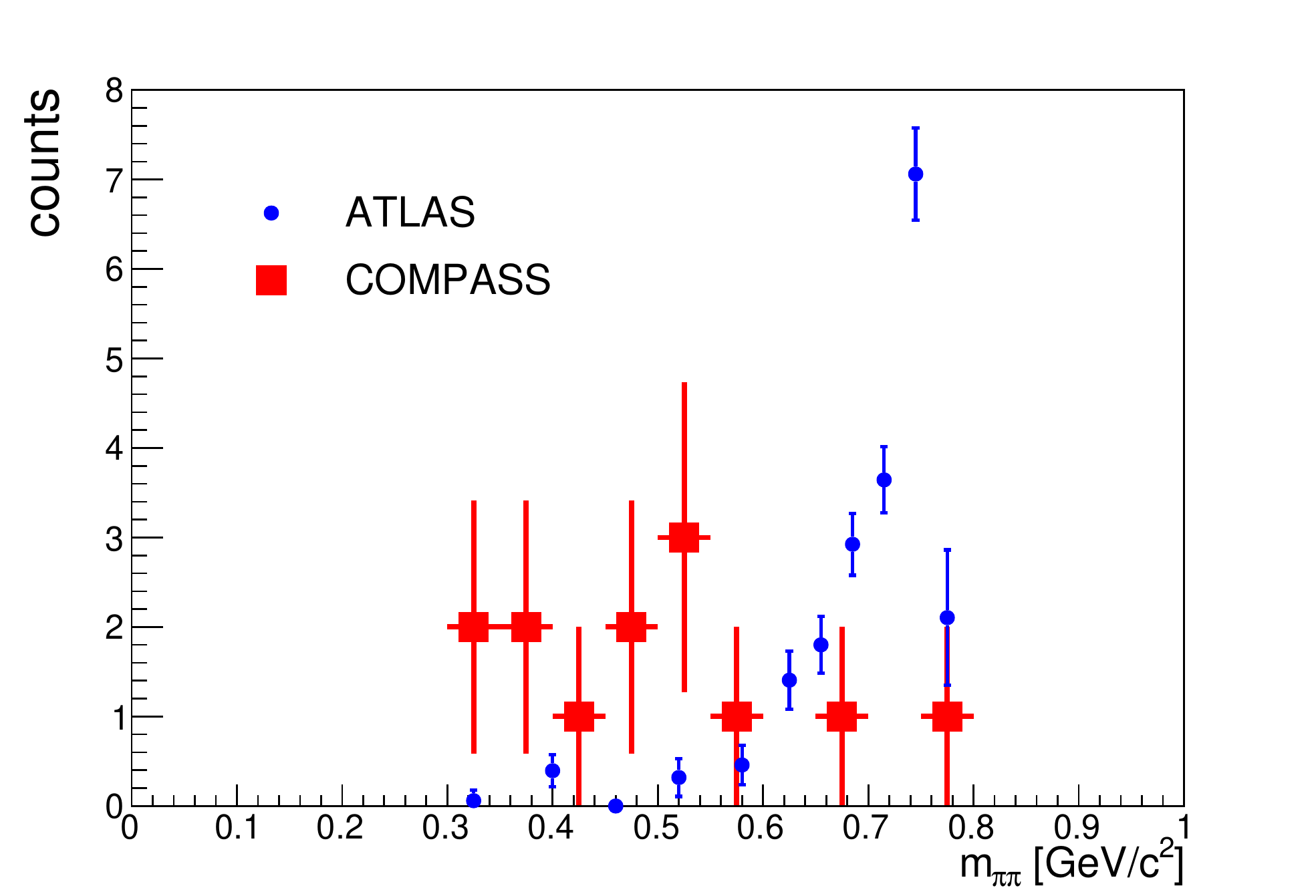}  
     \caption{\label{fig:pipi} Invariant mass spectra for the $\pi^+\pi^-$ subsystem from the decay of $\widetilde{X}􏰐(3872)$ measured by COMPASS  and from the decay of $X(3872)$ observed by ATLAS.}
\end{minipage}
 \end{figure}

Lepto(photo-)production of the $X(3872)$ at COMPASS was searched for in the exclusive charge-exchange reaction $\mu^+ N \rightarrow \mu^+ N' X(3872)\pi^{\pm}\rightarrow \mu^+ N' J/\psi \pi^+\pi^-\pi^{\pm}$. The invariant mass spectrum of the $J\psi\pi^+\pi^-$ subsystem is shown in Fig. \ref{fig:x3872} (two entries per event). It demonstrates two peaks 
whose positions and widths seem to suggest the production and decay of the $\psi(2S)$ and $X(3872)$ states. The statistical significance of the second peak is 4.1$\sigma$ that includes also systematic effects.
 Nevertheless, it was also found that the shape of the invariant mass distribution for $\pi^+\pi^-$  corresponding to the second peak looks very different from previous results obtained for $X(3872)$  by Belle, CDF, CMS and ATLAS (see Fig. \ref{fig:pipi}) and is inconsistent with quantum numbers 1$^{++}$.
 
We have to conclude that the observed signal is not the well-known X(3872) but a new charmonium state $\widetilde{X}(3872)$ that could be interpreted within the tetraquark model of Refs. \cite{Maiani:2004vq,Maiani:2014aja} which predicts a neutral partner of $X(3872)$ that has the similar mass, negative C-parity, and decays into $J/\psi\sigma$. The mass of the new resonance is estimated as 
$M_{\widetilde{X}(3872)}=3860.0\pm10.4$ MeV/$c^2$ and the Breit-Wigner width is $\Gamma_{\widetilde{X}(3872)}< 51$  
MeV/$c^2$ (CL=90\%). The  cross section of the reaction $\gamma N \rightarrow N' \widetilde{X}(3872)\pi^{\pm}$ multiplied by the branching fraction for the decay $\widetilde{X}(3872)\rightarrow J/\psi\pi^+\pi^-$ was found to be $71\pm28_{stat}\pm39_{syst}$ pb in the covered kinematic range with the mean value of $\sqrt{s_{\gamma N}}=14$ GeV.

\begin{figure}
 \begin{minipage}{16pc}
   \includegraphics[width=200px]{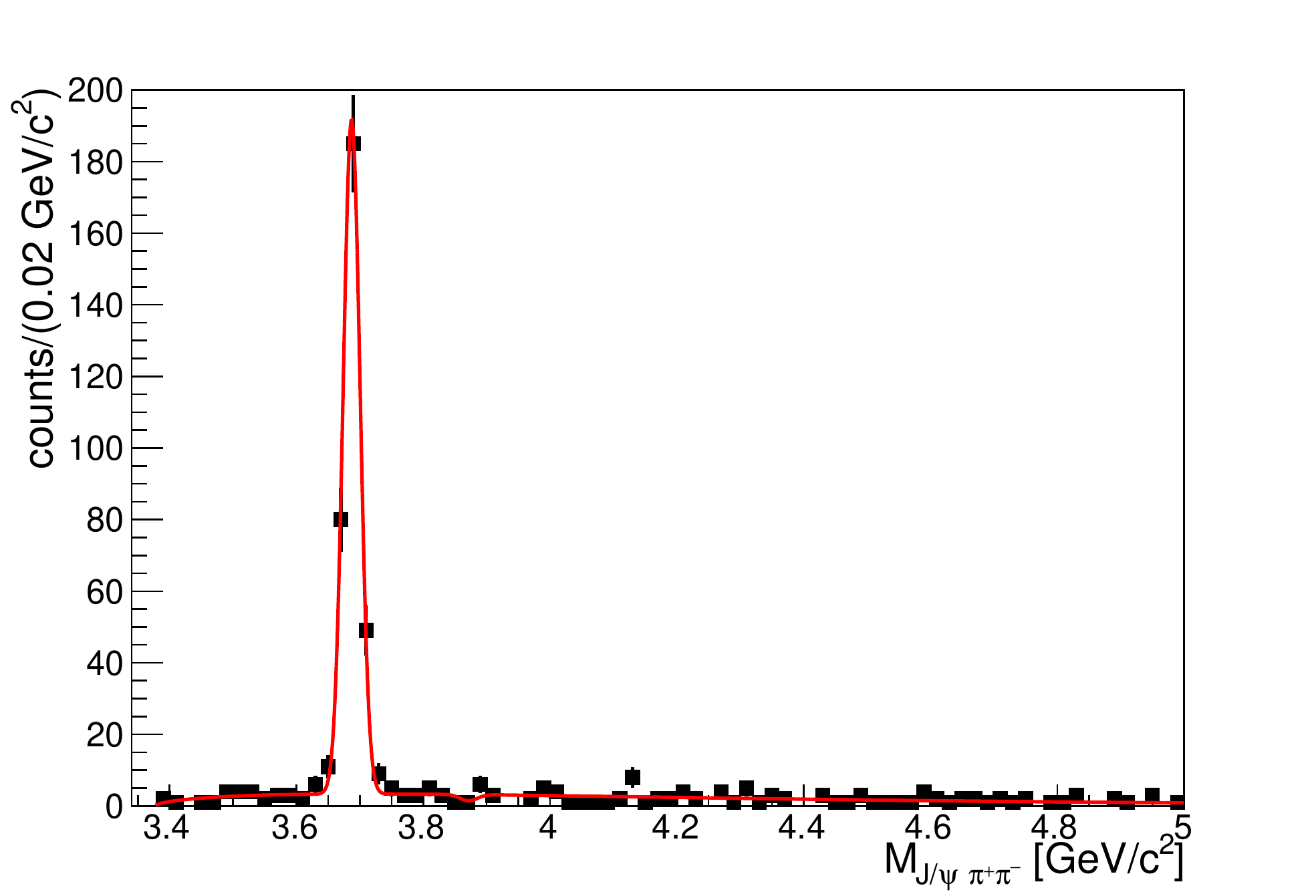}  
     \caption{\label{fig:x3872_neutral}
The $J/\psi\pi^+\pi^-$ invariant mass distribution for the final state in the reaction $\mu^+ N \to \mu^+ J/\psi\pi^+\pi^-N$. }
\end{minipage}\hspace{2pc}%
\begin{minipage}{20pc}
   \includegraphics[width=270px]{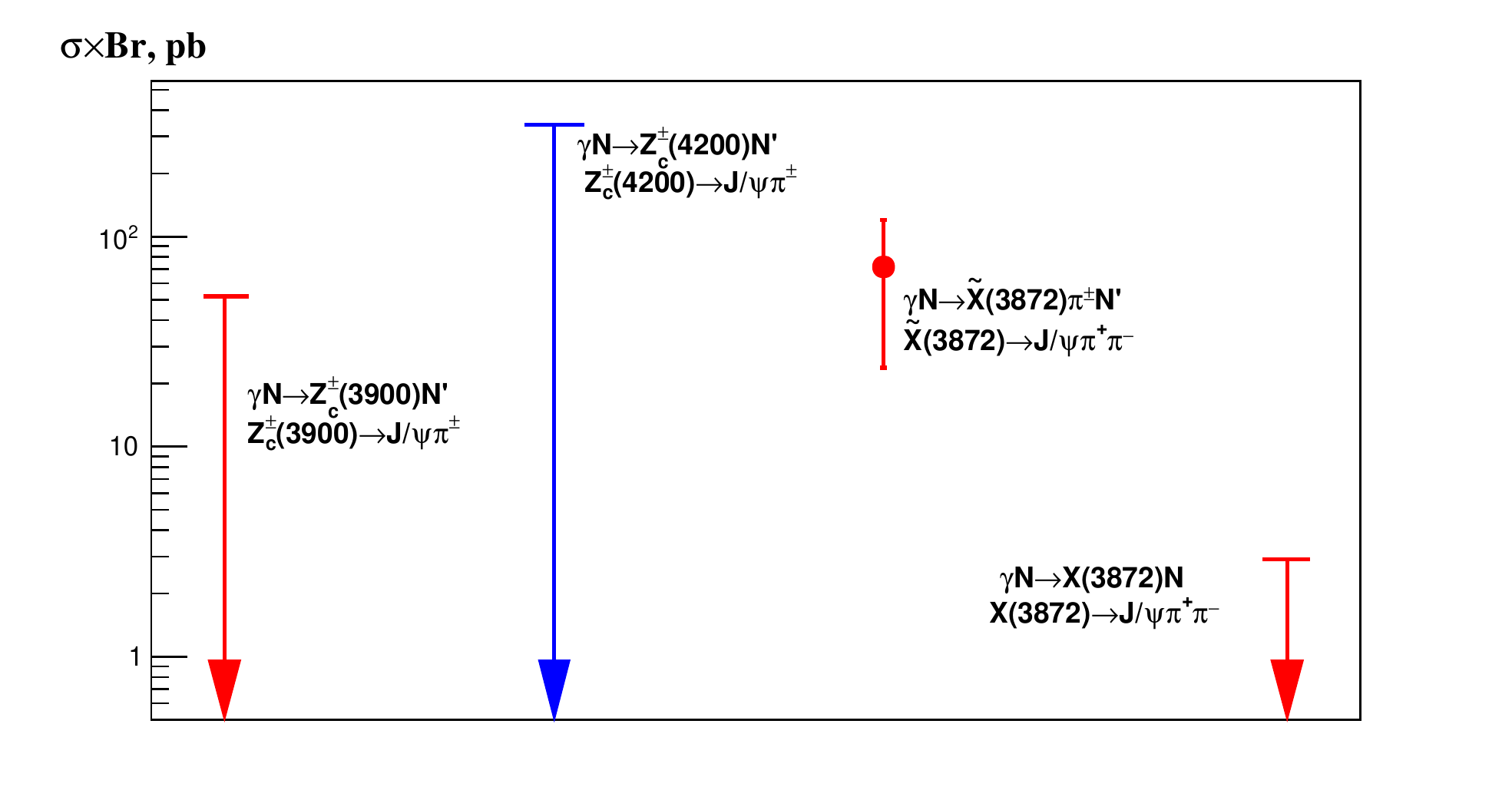}  
     \caption{\label{fig:total} Estimated rates of photoproduction of exotic charmonia obtained by COMPASS (red) or based on COMPASS results (blue).}
\end{minipage}
 \end{figure}

 Lepto(photo-)production of the $X(3872)$ was searched for also in the exclusive neutral reaction $\mu^+ N \rightarrow \mu^+ N' X(3872) \rightarrow \mu^+ N' J/\psi \pi^+\pi^-$. The invariant mass spectrum of the $J\psi\pi^+\pi^-$ final state is shown in Fig. \ref{fig:x3872_neutral}. It demonstrates only a peak from photo-diffractive production of $\psi(2S)$ while there is no statistically significant signal at around $M=3872$ MeV/$c^2$. The upper limit of the $X(3872)$ production cross section in this reaction multiplied by the branching fraction for the decay $X(3872)\rightarrow J/\psi\pi^+\pi^-$ was established to be below 2.9 pb (CL=90\%).
 
 More detailed information on the search for muoproduction of $X(3872)$ at COMPASS can be found in \cite{Aghasyan:2018vuq}.
 
\subsection{New possibilities}
The upgrade of the COMPASS setup related to the data taking in 2016--2017 within the framework of the GPD program \cite{proposal2} provides new opportunities to search for direct production of exotic charmonium-like states. A new, 2.5 m long liquid hydrogen target ($\sim$0.27$X_{0}$) is much more transparent for photons than the $^6$LiD and NH$_3$ targets that were used before. The 
target is surrounded by a 4 m long recoil proton detector which can be used to reconstruct and identify recoil protons via time-of-flight and energy loss measurements. The existing system of two electromagnetic calorimeters is extended by installation of the new large-aperture calorimeter. With the new calorimetry system one can expect much better selection of exclusive events. 

Searching for production of exotic charmonia that decay into final states with photons like $Z^0(3900)\to J/\psi\pi^0$, $X(3872)\to J/\psi \omega$, $\widetilde{X}(3872)\to J/\psi \eta$ etc. will be possible with the upgraded setup. The final states decaying to the $\chi_{c0,1,2}$ -mesons could also be studied.

New results on photoproduction of exotic charmonia could be expected from the LHC experiments where sizeable statistics of $J/\psi$, comparable with one available at COMPASS, produced in ultra-peripheral hadronic collisions is collected \cite{Aaij:2014iea,Abelev:2012ba,Kryshen:2017jfz,Khachatryan:2016qhq}. 
The Electron-Ion Collider  \cite{Accardi:2012qut} presently being discussed could also be an important source of new information about photoproduction of exotic charmonia.

\section{Conclusions}
Lepto(photo-)production of exotic charmonia is a new direction in physics of the XYZ states started by COMPASS. 
A new state $\widetilde{X}􏰐(3872)$ was observed  with the statistical significance of 4.1$\sigma$ in the exclusive charge-exchange reaction $\mu^+ N \rightarrow \mu^+ N' (J/\psi \pi^+\pi^-)\pi^{\pm}$. Its mass 
$M_{\widetilde{X}(3872)}=3860.0\pm10.4$ MeV/$c^2$, width $\Gamma_{\widetilde{X}(3872)}< 51$ and the decay mode $\widetilde{X}(3872)\to J/\psi\pi^+\pi^-$ are consistent with the X(3872). But the observed two-pion mass spectrum shows disagreement with previous experimental results for the X(3872). A possible explanation could be that the observed state $\widetilde{X}􏰐(3872)$ is the C = −1 partner of the X(3872) as predicted by a tetraquark model.

Estimated rates of photoproduction of exotic charmonia obtained by COMPASS or based on COMPASS results are summarized in Fig. \ref{fig:total} \cite{Zc,Z4200,Aghasyan:2018vuq}.

\end{document}